\documentclass[12pt,twoside,pointlessnumbers,smallheadings]{article}
\usepackage{graphicx}

\newcommand{\jpsi}{J/\psi}
\newcommand{\psip}{\psi(2S)}
\newcommand{\psipp}{\psi(3770)}
\newcommand{\rpi}{\rho\pi}
\newcommand{\ogpi}{\omega\pi^0}
\newcommand{\ddb}{D\bar{D}}
\newcommand{\EE}{e^+e^-}
\newcommand{\MM}{\mu^+\mu^-}

\newcommand{\PP}{\pi^+\pi^-}

\newcommand{\ra}{\rightarrow}
\newcommand{\beq}{\begin{equation}}
\newcommand{\edq}{\end{equation}}

\begin{document}
\parskip=5pt plus 1pt minus 1pt

\title{The Contribution of One-photon Annihilation at $\psip$ in $\EE$
Experiment \thanks{
Supported by National Natural Science Foundation of China (19991483)
and 100 Talents Program of CAS (U-25).}}

\author{{Wang Ping$^1$  Yuan Chang-Zheng$^1$  Mo Xiao-Hu$^{1,2}$ } \\
{\small 1 ( Institute of High Energy Physics, CAS, 
Beijing 100039, China ) ~~~~~~~~~~~~~~~} \\
{\small 2 ( China Center of Advanced Science and Technology,
Beijing 100080, China )} }
\date{\today}

\maketitle

\begin{center}
\begin{minipage}{15cm}
{\small {\bf Abstract} \hskip 0.5 cm
  The continuum one-photon annihilation at $\psip$ in 
  $\EE$ experiment is studied. Such contributions to the measured 
  final state $\ogpi$ and  $\PP$ at $\psip$ mass
  are estimated by phenomenological models. It is found that 
  these contributions must be taken into account in
  the determination of branching ratios of 
  $\psip\rightarrow  \ogpi$ and $\psip\rightarrow  \PP$, as
  well as other electromagnetic decay modes. 
  The study reaches the conclusion that in order for BES to 
  obtain the correct branching ratios on these decay modes, 
  at least 10pb$^{-1}$ of data below the $\psip$ peak is needed. \\

}
\end{minipage}
\end{center}

\section{Introduction}

The study and understanding of the decay dynamics of the charmonium states
$\jpsi$, $\psip$ and possibly $\psipp$, is one of the most important
topics in Quantum Chromodynamics (QCD). There have been lots of
activities in experimental part of this study in recent years since 
BES \cite{besii} has collected the world largest $\psip$ sample 
during 1993-1994 and 1995 running years at the $\psip$ peak.
Among the studies, the reconfirmation of the
vector-pseudoscalar (VP) decay puzzle (also called ``$\rpi$ puzzle'')
between $\jpsi$ and $\psip$ decays~\cite{rhopi} first observed by 
MARK II~\cite{mk2} and the first observation of 
the $\psip$ decay suppressions in vector-tensor (VT) 
modes~\cite{vt} compared to naive perturbative QCD predictions
are of great interest. Since then, many theoretical efforts have been 
put to figure out the possible reasons why $\jpsi$ and $\psip$ decays
to some final states, especially for VP and VT
modes significantly deviate from ``15 \% rule'', 
while many other decay modes agree 
with the predictions~\cite{rhopitheory}.

To extend the experimental study to other decay modes of $\psip$,
to further explore the studied modes to higher sensitivity, and to match
the huge sample BES has collected at the $\jpsi$ peak during 1999-2001
running years~\cite{bes2psp}, BES has collected 14~M $\psip$ data 
during 2001-2002 running year~\cite{bes2psp}. This world largest sample
makes it possible to measure the $\psip$ decays to electromagnetic 
decay level, so that one can reach the sensitivity of separating the QED
and QCD amplitudes in $\psip$ decays.

As for the study of $\psipp$, as an attempt to solve the ``$\rpi$ puzzle''
in $\psip$ decays, it has been proposed that the suppression of the 
$\psip$ decays to some specific final states is due to the mixing
between S- and D-wave charmonium states, or the $\psip$ is very close
to $\ddb$ threshold so that virtual process of $\psip \rightarrow \ddb$
cancels lots of $\psip$ decay amplitudes to those suppressed 
channels~\cite{rosner}. Furthermore, the study of the old data may
indicate $\psipp$ has a large fraction of decays into light 
hadrons besides $\ddb$ modes, in contradiction to the picture that 
it decays to $\ddb$ dominantly~\cite{ddbar}. 
To confirm this, the best way is to measure the $\ddb$ 
and total $\psipp$ cross sections with high precision, so that the difference
between the two gives the contribution of the non-$\ddb$ decay rate, 
while the other possible way is to measure the exclusive
non-$\ddb$ decay modes.

Now we know that $\jpsi$ and $\psip$ decay into hadrons, besides
those with charmonium in final particles, through two interactions:
the one-photon electromagnetic interaction and three-gluon strong interaction. 
The amplitudes of them, in general, may have a relative phase. This
is also true for $\psipp$ decaying into light hadrons.

In $e^+e^-$ colliding beam experiment, $J/\psi$ and $\psip$ are produced 
by $e^+e^-$ annihilation, there is inevitable the process
\begin{equation}
e^+e^- \rightarrow \gamma^* \rightarrow hadrons
\end{equation}
produced simultaneously, which is indistinguishable from the hadron events 
from $J/\psi$ or $\psip$ decays. So in $e^+e^-$ experiment,
any final state may come from three processes: the charmonium
three-gluon decays, the charmonium one-photon decays, and the 
virtual photon process without going through resonance (continuum process). 
The study of the charmonium decay dynamics is to determine 
the decay amplitudes of the three-gluon and the one-photon processes.
In the case of the $\psip$, the one-photon decays 
has a comparable cross section with respected to the continuum process.
Moreover, for some strongly suppessed modes like VP and VT,
the three-gluon process may also have a comparable cross 
section. In this case, one has to consider three amplitudes 
and the relative phases between any of the two amplitudes in the
analysis of the experimental data.
Fortunately, there are some decay modes of the charmonium 
where the strong decays are forbidden, only 
decays through one-photon annihilation is allowed, 
like $\psip$ $\rightarrow\ogpi$, which
violates the isospin, and $\psip$ $\rightarrow\pi^+\pi^-$,
which violates the G-parity. In such situation, only two amplitudes 
and one relative phase are present, this 
substantially simplifies the study, and makes the
determination of the one-photon decay amplitude possible, 
provided the amplitude of the continuum process being known.

\begin{figure}[hbt]
\begin{minipage}{7.5cm}
\includegraphics[width=7. cm,height=5.5cm]{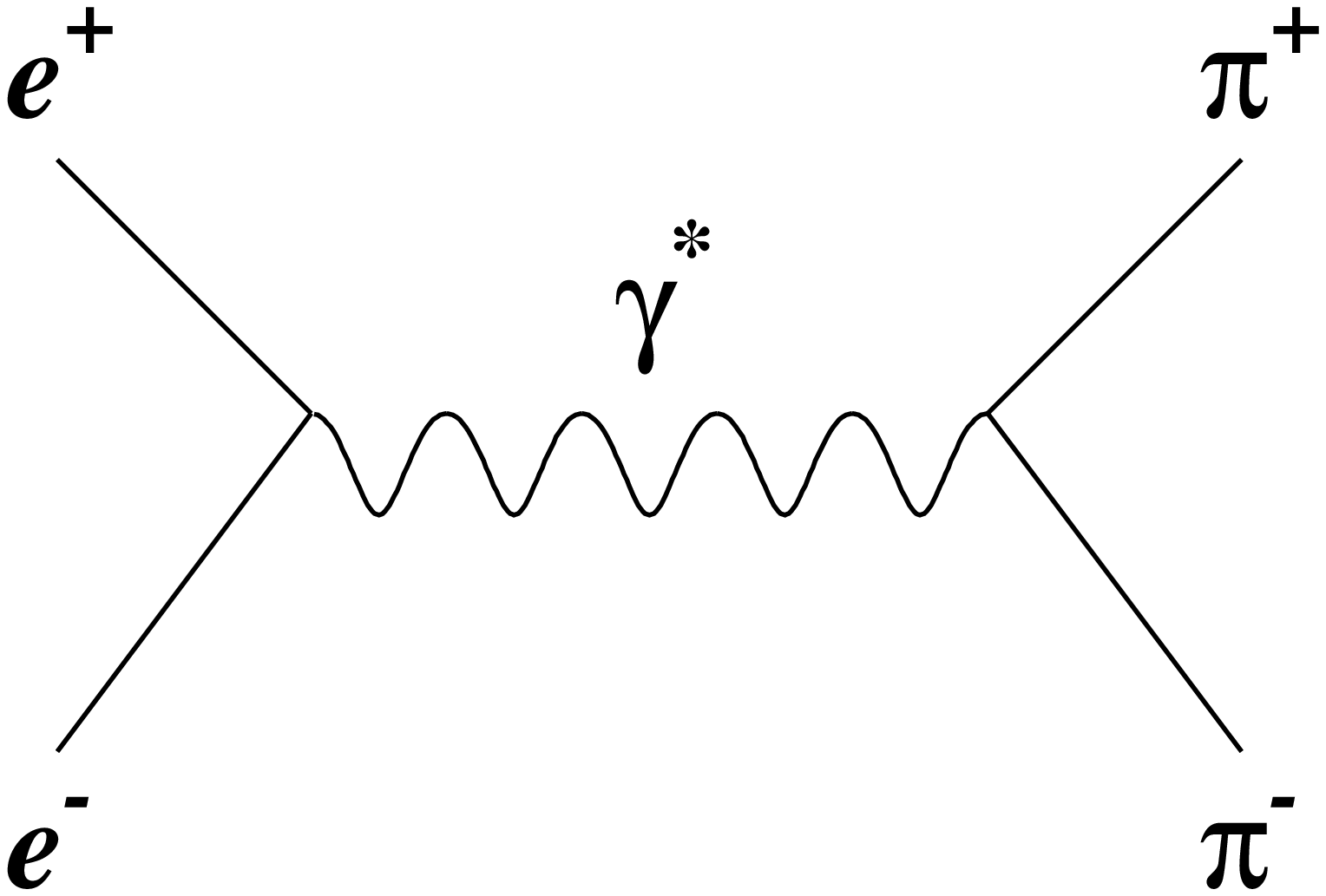}
\end{minipage}
\hskip 1.5cm
\begin{minipage}{7.5cm}
\includegraphics[width=7.cm,height=5.5cm]{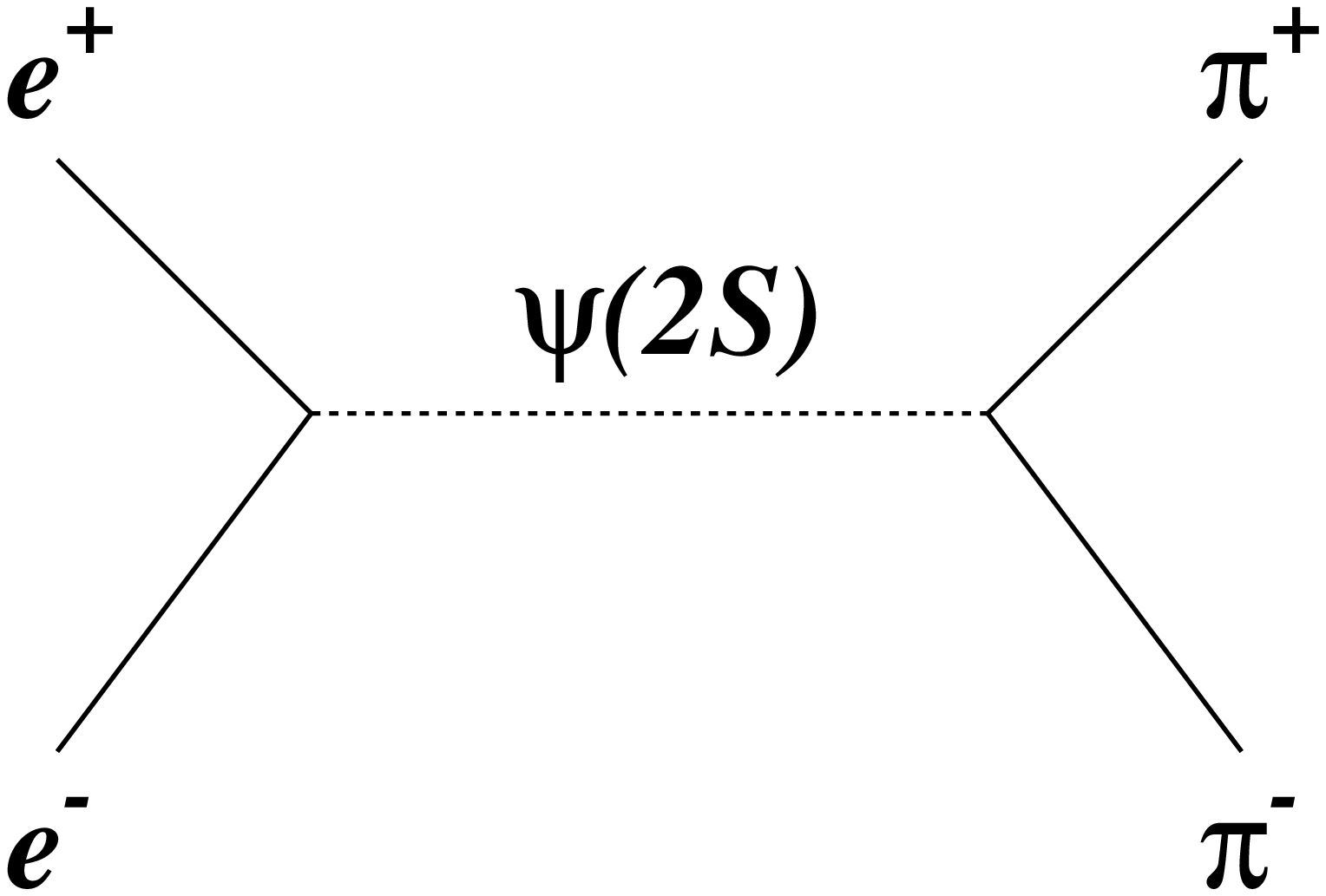}
\end{minipage}
\caption{\label{twopros} Two Feynman diagrams of $\EE \rightarrow \PP$ 
final state at $\psip$.}
\end{figure}

The electromagnetic processes, such as $\EE \rightarrow \ogpi$
and $\EE \rightarrow \PP$, are similar to the 
process $e^+e^- \rightarrow \mu^+\mu^-$ in the way that there are
two Feynman diagrams : the continuum one-photon diagram 
and the $\psip$ diagram. Taking $\PP$ final state as an example, two
diagrams are shown in Fig. \ref{twopros}. 

The observed experimental cross sections of these 
exclusive channels consist of three parts:
the $\psip$ resonance, the continuum and their interference. 
However, unlike the $\MM$ channel which the continuum
amplitude can be calculated by QED, such terms for $\ogpi$ 
and $\pi^+\pi^-$ are to be determined by experiment, i.e., we need 
to measure the form factors of $\omega\pi^0$ and $\pi^+\pi^-$.

It is important to notice that the measured cross sections 
at the narrow resonances are sensitive to the experimental conditions, 
but the three parts in the cross section depend on different aspects
of the experimental details. The measured resonance cross
section depends on the energy spread of  the $e^+e^-$ collider,
but such finite energy spread does not affect the continuum term.
On the other hand, the continuum term is sensitive
to the invariant mass cut in event selection, while such cut
hardly affects the observed resonance cross section under practical 
event selection criteria. These are to be discussed in detail in 
the following sections.

The study of the pure electromagnetic decays of the charmonium
states will shed light on the understanding of the charmonium decay
dynamics. Some theorists, like Suzuki~\cite{suzuki},
have tried to probe the interference pattern between one-photon 
electromagnetic process and three-gluon process of the charmonium
decays in order to 
solve the ``$\rho\pi$ puzzle''. For such analysis, the
experimental information of those decays of pure electromagnetic
processes, are of particular importance~\cite{chernyak}, since 
they supply an estimation of the electromagnetic part in the 
decay modes with strong interaction.  

In this paper, we first study the experiment dependence 
of the cross sections, then give an estimation of the cross sections 
of the pure electromagnetic process without going through charmonium
resonance based on the form factors, finally the minimum required
integrated luminosity at off-peak energy point is estimated if 
meaningful results are expected for the decay modes interested.

\section{Experimentally observed cross sections}

\subsection{Resonance}

The cross section of the process 
$e^{+} e^{-} \rightarrow \psip \rightarrow f $
(where $f$ denotes a certain kind of final state) is described
by the Breit-Wigner formula
\begin{equation}
\sigma_{BW}(W)=\frac{12 \pi \cdot \Gamma_{e} \Gamma_{f} }
                   {(W^2-M^2)^2+\Gamma_t^2 M^2} ,
\label{breit}
\end{equation}		   
where $W$ is the center-of-mass energy, 
$\Gamma_{e}$ and $\Gamma_{f}$ are the widths of 
$\psip$ decaying into $e^{+} e^{-}$ and $f$, $\Gamma_t$ and $M$
are the total width and mass of $\psip$. Taking the initial state 
radiative correction into consideration, the cross section 
becomes \cite{rad.1} 
\begin{equation}
\sigma_{r.c.} (W)=\int \limits_{0}^{x_m} dx 
F(x,s) \frac{1}{|1-\Pi (s(1-x))|^2} 
\sigma_{BW}(s(1-x)) ,
\end{equation}    
where $s=W^2$, $x_m=1-s'/s$,
$\sqrt{s'}$ is the experimentally required minimum invariant 
mass of the 
final state $f$ after losing energy due to multi-photon emission; 
$F(x,s)$ has been calculated in many references \cite{rad.1, rad.2, rad.3}
and $\Pi (s(1-x))$ is the vacuum polarization factor.
The radiative correction in the final states are usually
not considered~\cite{tsai,barends}. The reasons are twofold. 
In the first place, the hadronic final system is very 
complicated and since the radiative corrections depend
upon the details of how the experiment is done, it is difficult
to give a general, model-independent prescription for them. The 
second reason is that our understanding of the hadronic problem
is so crude that there is no need to worry about the 
electromagnetic corrections. In any case, if we find later 
on that it is necessary to do radiative corrections to the 
hadronic states for some specific problem, we can do the 
calculation then, because the initial state radiative 
corrections and final state radiative corrections can be 
decoupled to a large extent. 

The $e^+e^-$ colliders have finite energy spread. 
The energy spread function $G(W,W')$ is usually 
a Gaussian distribution :
\begin{equation}
G(W,W^{\prime})=\frac{1}{\sqrt{2 \pi} \Delta}
             e^{ -\frac{(W-W^{\prime})^2}{2 {\Delta}^2} },
\end{equation}	     
where $\Delta$ is the standard deviation of the Gaussian 
distribution. It varies with the beam energy of the collider. 
In case of BEPC/BES, $\Delta=1.3$ MeV at the C.M. energy of the 
$\psip$ \cite{delta}. It is much 
wider than the $\psip$ intrinsic width
of (300 $\pm$ 25) keV~\cite{PDG}.
So the experimentally measured resonance cross section is the
radiatively corrected Breit-Wigner cross section folded with the
energy spread function: 
\begin{equation}
\sigma_{exp} (W)=\int \limits_{0}^{\infty}
        dW^{\prime} \sigma_{r.c.} (W^{\prime}) G(W^{\prime},W),
\end{equation}
where $\sigma_{r.c.}$ is defined by Eq.(3).
\begin{figure}[hbt]
\begin{center}
\includegraphics[width=10.cm,height=8.cm]{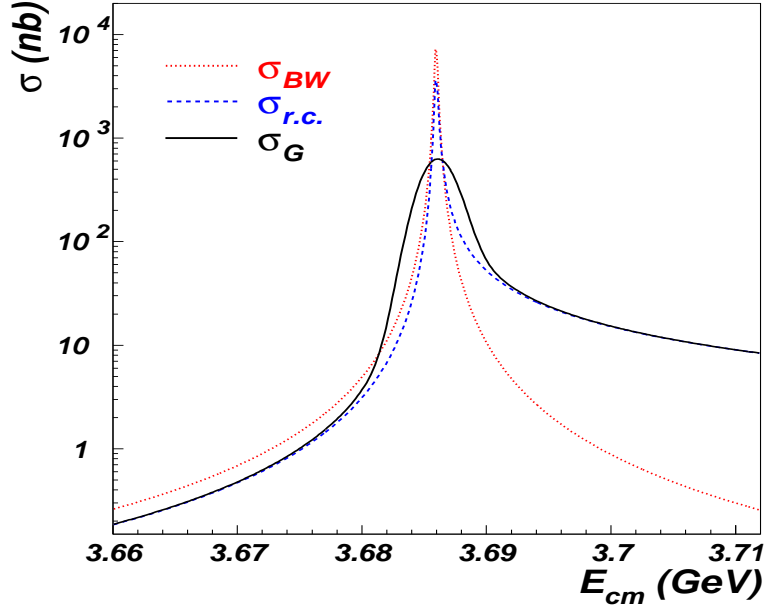}
\caption{\label{cmpbrg} Inclusive hadronic cross section of $\psip$: 
$\sigma_{BW}$ for Breit-Wigner cross section; 
$\sigma_{r.c.}$ the cross section with radiative correction; 
$\sigma_{exp}$ the measured cross section on a collider with $\Delta=1.3$ MeV.}
\end{center}
\end{figure}

In Fig. \ref{cmpbrg}, three cross sections are depicted: the 
Breit-Wigner cross section of Eq.(2); the cross section after
radiative correction by Eq.(3), and the experimentally measured 
cross section by a $e^+e^-$ collider with $\Delta=1.3$ MeV. 
In the calculation of these cross sections, 
the following parameters of $\psip$
are used~\cite{PDG} : $M=3.68596$ GeV; $\Gamma_{t}=300$ keV;
$\Gamma_{e}=2.19$ keV~\footnote{Throughout this paper, we 
use this set of data and  $\Delta=1.3$ MeV for numerical 
calculations.}. From the three curves in Fig.~\ref{cmpbrg},
we see that the radiative correction reduces
the height of the resonance. It also shifts the peak position
to above the $\psip$ nominal mass.
The reduction factor $\rho$ and the shift of the peak 
$\Delta\sqrt{s}_{max}$ are approximately expressed by~\cite{berends}

\begin{eqnarray}
\rho &=&{\displaystyle \left(\frac{\Gamma_{t}}{M}\right)^\beta
 \cdot (1+ \delta)}, \\
\Delta \sqrt{s}_{max} &=&{\displaystyle \frac{\beta \pi}{8} \Gamma_{t} },
\end{eqnarray}
where $\beta$ is defined as
$$ \beta = \frac{2 \alpha}{\pi} \left( \ln \frac{s}{m^2_e} -1 \right), $$
and 
\begin{equation}
\delta = \frac{3}{4}\beta+\frac{\alpha}{\pi}
\left(\frac{\pi^{2}}{3}-\frac{1}{2}\right)+\beta^{2}
\left(\frac{9}{32}-\frac{\pi^{2}}{12}\right) .
\end{equation}

At the $\psip$ mass, $\beta \approx 0.0779$
and $\delta \approx 0.06$. So for $\psip$, the 
reduction factor $\rho \approx 0.51$  and the  shift of the peak 
$\Delta \sqrt{s}_{max} \approx 9$ keV.
The energy spread further lowers down and shifts 
the experimentally measured $\psip$ peak. In the case of a collider 
with $\Delta=1.3$ MeV, the maximum height of the $\psip$ peak becomes 640 nb, 
and the position of the peak is shifted by 0.14MeV above the $\psip$
nominal mass.

\begin{figure}[hbtp]
\begin{center}
\begin{minipage}{12.cm}
\includegraphics[width=12.cm,height=8.cm]{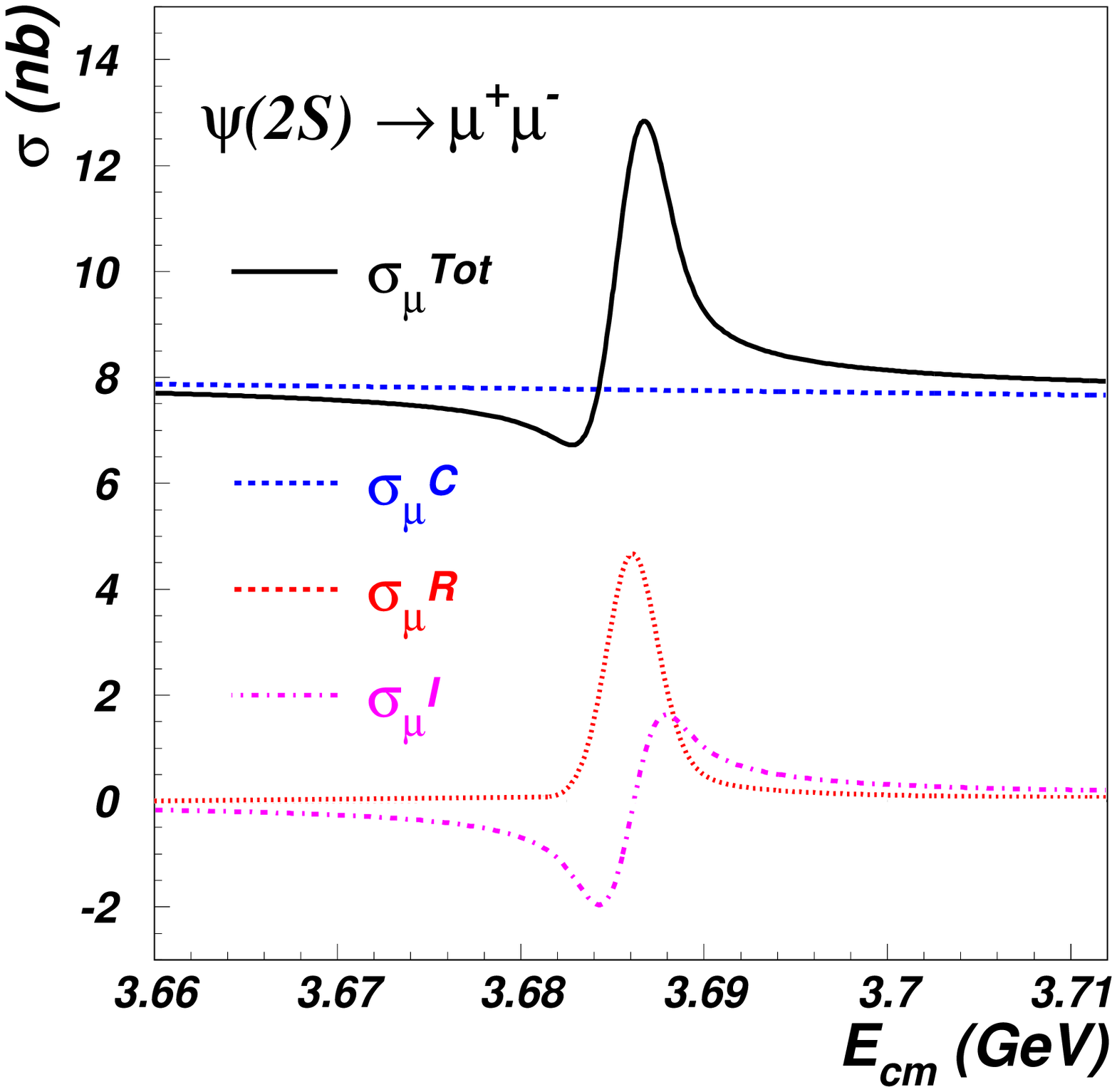}
\caption{\label{cmpuirc} Contributions of three parts to the cross section: 
dashed line for QED continuum ($\sigma^C$); 
dotted line for resonance ($\sigma^R$);
dash dotted line for interference($\sigma^I$); 
solid line for total cross section($\sigma^{Tot}$).}
\end{minipage}
\vskip 0.5cm
\begin{minipage}{12.cm}
\includegraphics[width=12.cm,height=8.cm]{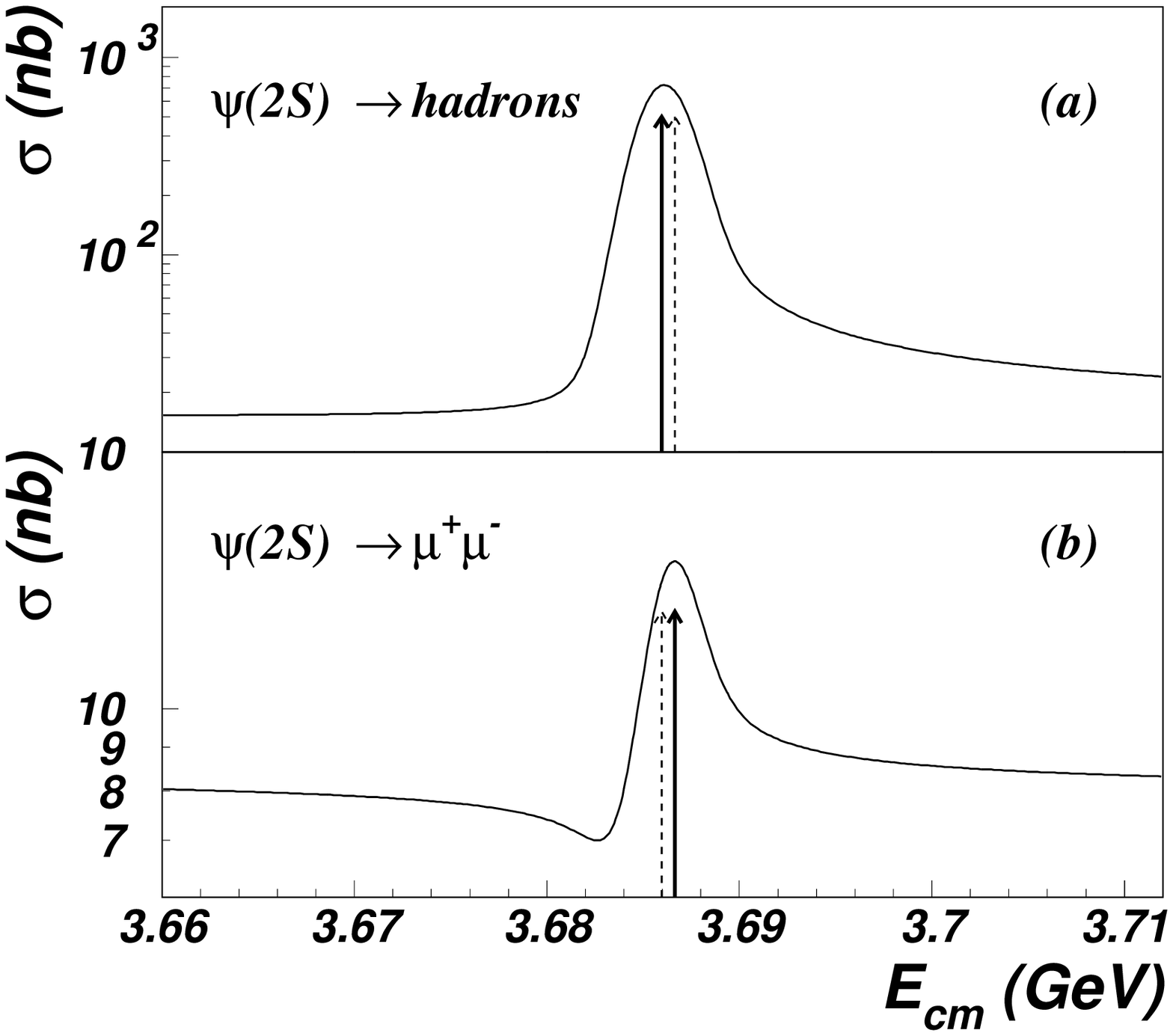}
\caption{\label{cmphup} Cross sections in the vicinity of $\psip$
for hadron (a) and $\MM$ (b) final states.
The solid line with arrow indicates the peak position.} 
\end{minipage}
\end{center}
\end{figure}
  
The $\MM$ channel deserves special discussion here, 
since it is parallel to those hadronic channels in $\psip$ decays  
which only go through electromagnetic interaction, such as
$\ogpi$ and $\pi^+\pi^-$.
Since this is an exclusive channel, there is interference between
the continuum and the $\psip$ amplitudes.
Such interference can be seen clearly from the scan of the $\psip$,
see Fig. \ref{cmpuirc}. In Fig. \ref{cmphup}, the cross sections of 
inclusive hadrons and $\mu^+\mu^-$ 
pairs are depicted for comparison. 
Here in the calculation of radiative correction, 
the upper limit of integration $x_m$ in Eq.(3) is taken to be
($1-4m_{\mu}^2/s$), which means all allowed phase space
for multi-photon emission is integrated.  The peak of $\mu^+\mu^-$ curve 
is shifted more than that of the inclusive hadrons, 
to 0.81 MeV above the $\psip$ nominal mass. 

\subsection{One-photon continuum}

In the total experimentally measured cross section, the 
one-photon continuum term has different features
from the resonance. For the $\MM$ channel, 
this term is expressed in the Born order as
\beq
\sigma_{\gamma^* Born}(s) = \frac{4\pi\alpha^2}{3s} .
\edq
After radiative correction, this term depends sensitively on the 
upper limit of the integration $x_m$ in Eq.(3). At the C.M. energy 
of the $\psip$ mass, the radiatively corrected cross section is 8.33 nb,
when $x_m=1-4m_{\mu}^2/s$. This value means integrating 
all the allowed phase space of multi-photon emission.
If $x_m=0.2$, which means that the final $\MM$ has
the minimum invariant mass of $\sqrt{(1-0.2)}M_{\psip}$,
after losing energy to multi-photon emission, then the radiatively 
corrected cross section is 6.25 nb. In the actual experimental situation, 
the invariant mass cut or its equivalence is usually 
imposed to remove the $\MM$ from $J/\psi$ decays. 
For the resonance term, as long as 
\beq
 x_m \gg \frac{\Gamma_{t}}{M},
\edq
it is insensitive to $x_m$. Another important feature 
of the continuum term is that since it is a smooth function 
of $s$, the finite energy spread of the $\EE$ collider does not 
change the measured value. 

For the inclusive hadronic final states, the continuum
term is expressed in the Born order as
\beq
\sigma_{\gamma^* Born}(s) = \frac{4\pi\alpha^2}{3s}R(s),
\edq
where $R(s)$ is the $R$-value at C.M. energy $\sqrt{s}$. 
Since $R(s)$ is a slowly varying function of energy, so 
qualitatively it shares the same feature of this term for $\MM$
final state. After radiative correction,
it depends sensitively on the upper limit
of the integration $x_m$ in Eq.(3), i.e., on the experimental cuts;
but the finite energy spread of the $\EE$ collider hardly 
changes the measured value.

\section{Cross sections of two exclusive electromagnetic processes}

The general discussions in previous section is extended to 
exclusive processes. For illustrative purpose, we discuss two pure 
electromagnetic ones, which are similar to $\EE\ra \MM$.

\subsection{Resonance}

In experiments, usually the branching ratio is measured by
the events number of certain channel ($N_f$) divided by the total number of
the resonance events($N_{\psip}$):
\beq
{\cal B}_f  = \frac{N_f}{N_{\psip}} = \frac{\sigma_{f}}{\sigma_{\psip}} ,
\label{brsgsg}
\edq
where $\sigma_{f}$ is the measured cross section of the specific
channel and $\sigma_{\psip}$ the total production cross section of $\psip$.
With the $\psip$ parameters and $\EE$ collider energy 
spread used in previous section, the total cross section 
of $\psip$ is 640 nb (=$\sigma_{\psip}$). The decay 
$\psip\rightarrow \omega\pi^0$ is reported a branching ratio of
$(3.6\pm1.6\pm0.6) \times 10^{-5}$~\cite{omegapi}; the other decay
$\psip\rightarrow \pi^+\pi^-$ is reported a branching ratio of
$(8\pm5)\times10^{-5}$ by DASP \cite{dasp}, but a later result by BES is
$(8.4\pm3.5^{+1.6}_{-1.3})\times10^{-6}$~\cite{pipi}. 
If we check the original references, 
these numbers actually mean that 
the BES measured cross section of $e^+e^-\rightarrow \ogpi$ 
at $\psip$ mass is $2\times10^{-2}$nb while for 
$e^+e^-\rightarrow \pi^+\pi^-$ it is $5\times10^{-3}$nb.

We notice that the observed cross section of 
these final states are 3 to 4 orders of
magnitude smaller than the total inclusive hadron cross section of the
continuum process which, according to Ref. \cite{scanpsip},
is about 15 nb. So it could be that a substantial part of the 
experimentally measured cross section comes from the continuum
instead of the $\psip$ decay.  Therefore it is 
essential to know the production rate
of $\ogpi$ and $\pi^+\pi^-$ due to the continuum process 
in order to get the correct 
branching ratios of $\psip$ decays to these modes.

\subsection{One photon continuum}

Here we give an estimation of the possible magnitudes of 
$\EE \rightarrow \gamma^{\ast} \rightarrow \ogpi$ and
$\EE \rightarrow \gamma^{\ast} \rightarrow \PP$
by currently available phenomenological models. These two
processes are calculated by their form factors
\begin{eqnarray}
\sigma_{\omega \pi}(s) &=&{\displaystyle 
\frac{4 \pi \alpha^2}{3} \cdot \frac{|p|^3}{s^{3/2}} \cdot
| F_{\omega \pi} (s) |^2 },\\
\sigma_{\pi \pi}(s)    &=&{\displaystyle 
\frac{8 \pi \alpha^2}{3} \cdot \frac{|p|^3}{s^{5/2}} \cdot
| F_{\pi \pi} (s) |^2 },
\end{eqnarray}
where $p$ is the momentum of the final particles, 
$F_{\omega \pi} (s)$ and $F_{\pi \pi} (s)$
are the form factors for $\omega \pi$ and $\PP$ respectively.
Phenomenological model~\cite{chernyak} gives
\begin{equation}
  F_{\omega\pi}(s)=F_{\omega\pi}(0)\frac{m_{\rho}^2M_{\rho'}^2}
              {(m_{\rho}^2-s)(M_{\rho'}^2-s)} ,
\label{facogpi}
\end{equation}
with
\begin{equation}
F_{\omega\pi}(0)=2.3 \mbox{  GeV$^{-1}$} .
\end{equation}
At $E_{cm}=3.686$ GeV, it gives $3.6\times10^{-3}$nb at Born order.  
It is almost unchanged by radiative correction if 
$x_m=0.2$ in Eq.(3)\footnote{The radiatively corrected cross
section is $3.9\times10^{-3}$nb 
if $x_m=0.3$ and $3.3\times10^{-3}$ if $x_m=0.1$.}.

For $\pi^+\pi^-$, the
vector dominance model (VDM)~\cite{achasov} gives :
\begin{equation}
F_{\pi\pi}^{VDM}(S)=\frac{m_{\rho}^2}{m_{\rho}^2-s} .
\end{equation}
At $\psip$ mass, it gives $3.2\times10^{-3}$nb at Born order, which
is almost unchanged after radiative correction if $x_m=0.2$\footnote{The 
radiatively corrected cross section   
is $3.4\times10^{-3}$nb if $x_m=0.3$ and $3.0\times10^{-3}$nb 
if $x_m=0.1$.}. 

From above estimations, we can see that possibly some of the 
observed $\ogpi$ and a large fraction 
of the observed $\pi^+\pi^-$ events at $\psip$ peak may come
from the continuum process instead of the $\psip$ decay. 
It should be noted that different models for the form factors
give very different results at the $\psip$ mass
region~\cite{achasov}, here we only want to get a feeling 
about the magnitude of the continuum contribution.

For $\psipp$, the resonance cross section is only around 8 nb; 
whereas the hadron cross section from the continuum
is about 13 nb. To determine its branching ratio of exclusive light hadrons, 
we must know the continuum cross section of these final states.

\setcounter{footnote}{0}
\subsection{Energy dependence of measurement}

From the foregoing discussion, we know that the behavior of two 
terms of the cross section is rather distinctive with the energy 
variation: the resonance changes manifestly but the continuum
part almost imperceptibly. The relative proportion of two terms of 
the cross section, which plays an important role in determination 
of branching ratio, could be rather different at different energy.

In actual experiment, the beam energy could drift away
from the position of the maximum inclusive hadron cross section 
during a considerable long time running, which in turn
leads to considerable variation of resonance cross section.
Taking $\MM$ final state as an example, if the energy
drifts upwards or downwards from the $\psip$ peak position of 
the inclusive hadron by 0.5 MeV, the $\MM$ resonance cross section 
changes from 4.47 to 4.20 nb,
or equivalently the variation is about 6\%.
 
Furthermore, for the process whose continuum cross section is 
comparable with that of resonance, its interference cross section accounts
for a considerable part in the total cross section, and varies violently 
in the vicinity of the peak position. Also for $\MM$ final state,
its interference cross section changes from $-0.14$ nb to $-0.96$
or 0.66 nb respectively for the energy drifted by 0.5 MeV lower
or higher than the hadron peak position.

Combining the variation of the resonance and the interference, 
the maximum change of the resonance cross section can reach 1.1 nb, 
or 25\% for $\MM$ channel.
In other words, if we use the result from $\MM$
final state as an approximation, 25\% uncertainty should be taken 
into account in the branching ratio estimation 
for the 0.5 MeV energy drift\footnote{If the energy drift is about
0.2 MeV, the uncertainty is only 9\% correspondingly.}.

\section{Other $\psip$ decay modes}

In previous sections, we have discussed the pure electromagnetic processes
in $\psip$ decays. Most other $\psip$ decays which strong 
interaction plays the leading role also have continuum
amplitude contribution if the $\psip$ is produced in $\EE$ 
collision. Although the amplitude maybe smaller, but for those 
suppressed channels, like VP and VT modes, the 
electromagnetic decays and the continuum contribution 
could play an important role.
For example, now the upper limit of $\psip \rightarrow \rho\pi$
branching ratio has been pushed down to $2.9\times 10^{-5}$~\cite{rhopi}. 
Under BES condition, this means that the upper limit 
of the cross section is $2\times10^{-2}$ nb. The 
coupling of this channel to virtual photon is one third of 
$\ogpi$. Using the estimate of previous sections, its 
continuum cross section could be $1.2\times10^{-3}$nb at $\psip$.
If we push down further the upper limit of this decay, we  
need to take into account the contribution from the continuum. 
This is also true for
the upper limit of $\psip \rightarrow K^+K^*(892)^- +c.c.$.
Another VP decay mode,
$\psip \rightarrow K^{*0}\bar{K^0}+c.c.$, has been measured 
to have a branching ratio of 
$(0.81\pm0.24\pm0.16)\times10^{-4}$~\cite{kstark}. This means 
that BES measured cross section of this channel is
$5\times10^{-2}$nb at $\psip$. Theoretically, the coupling of this channel
to virtual photon is two thirds of $\ogpi$, so the
estimated continuum cross section by the form factors of 
Eq. (\ref{facogpi}) is $3.2\times10^{-3}$nb. This 
has to be considered in high precision experiments of the coming generation 
accelerator and spectrometer, like CLEO-c and BES-III.

In order to know whether the observed  
suppression of VP and VT modes in $\psip$ decays 
are due to the absence of strong interaction
amplitude, or the destructive interference between the electromagnetic
and the strong amplitudes, or just an incidental destructive
interference between these two and the continuum process
in our particular experiment, we need to know their coupling to 
virtual photon.
 
\section{Estimation of needed data taking at BES}

As discussed in previous sections, in order to extract the branching
ratios of $\psip$ to $\ogpi$ and $\PP$ from experimental data, 
we need the coupling of these channels to a virtual photon.
This can only be done by 
the direct measurement of their production cross sections nearby but 
off $\psip$ resonance. The same is true for $\psipp$ decaying into 
light hadrons. 
To satisfy the need of both $\psip$ and $\psipp$ physics, the best
energy to take the data should be below $\psip$ peak, for example,
at $E_{cm}=3.67$ GeV. 
Here two factors should be considered: one is the data-taking 
point should not be too faraway from the peak position, so that the 
cross section of the continuum process is almost equal to that at the peak 
position; another factor is the data-taking point could not be
too close to the peak position, so that the resonance cross section 
is small enough to be neglected. Taking the inclusive hadron final
state as an example, at $E_{cm}=3.67$ GeV, the cross section 
variation of the continuum process 
\beq
\delta_{\sigma_{\gamma^{\ast}}}=
\frac{\sigma_h(\mbox{$E_{cm}=3.67$ GeV})-\sigma_h(\mbox{$E_{cm}=3.686$ GeV})}
{\sigma_h(\mbox{$E_{cm}=3.67$ GeV})} \approx 1.05 \% .
\edq
At the same time, the ratio of the resonance cross section to that of 
the continuum process
\beq
R \left(\frac{\mbox{resonance}}{\mbox{continuum}} 
  \right)_{\mbox{$E_{cm}=3.67$ GeV}}=
 \left( \frac{\sigma_{\EE \rightarrow \psip \rightarrow hadron} }
     {\sigma_{\EE \rightarrow \gamma^{\ast} \rightarrow hadron} }
 \right)_{\mbox{$E_{cm}=3.67$ GeV}} \approx 2.18 \% .
\edq
That is to say, at $E_{cm}=3.67$ GeV, the variation of the continuum process
cross section and the proportion of the resonance cross section are both
fairly small.

Another reason to take the data below $\psip$ peak is
that $\psip$ and $\psipp$ are too close to find a suitable data-taking
point between them. On one hand, $\psipp$ is a wide resonance,
the resonance effect can extend far beyond the peak position. On the
other hand, the effect from the radiative tail of $\psip$ can reach
the vicinity of the $\psipp$ peak. For example, 
at one $\Gamma(\psipp)$ below $\psipp$ peak, 
the resonance cross section is $0.8$ nb, which is mostly $D\bar{D}$; 
the radiative tail of $\psip$ is more than 3 nb; 
while the cross section for continuum process is about $13$ nb.
That is to say, the radiative tail of $\psip$ is too large to be neglected.

Finally, we discuss quantitatively how much data are needed.
\begin{itemize}
\item The typical electromagnetic process, like $\psip \rightarrow
\omega\pi^0$, is measured to have a branching ratio of
$3\times10^{-5}$, since the
cross section by BEPC/BES at $\psip$ peak is 640 nb, that means the
$\omega\pi^0$ cross section is $2\times10^{-2}$ nb. 
So for every pb$^{-1}$, we obtain 20 produced events. 

If we use the form factor in Eq.(13) to estimate the cross section, at  
3.67GeV it is $3.8\times10^{-3}$ nb with $x_m=0.2$ in radiative correction. 
If so, for every pb$^{-1}$, we obtain 3.8 produced events.  

\item Another pure electromagnetic process $\psip \rightarrow \pi^+\pi^-$
is measured to have a branching ratio of $8\times10^{-6}$, 
corresponding cross section is about $5\times10^{-3}$ nb.
For every pb$^{-1}$, we obtain 5 produced events.

If we use the form factor in Eq.(14) to estimate the cross section, at  
3.67GeV it is $3.3\times10^{-3}$nb with $x_m=0.2$ in radiative correction. 
If so, for every pb$^{-1}$, we obtain 3.3 produced events.  

\item $\EE \rightarrow \rho\pi$ is one of the most interesting processes.
Each of the three final states, $\rho^0\pi^0$, $\rho^+\pi^-$, $\rho^-\pi^+$
gets $1/9$ contribution as $\ogpi$ from pure one-photon
process. So for every pb$^{-1}$, we expect 6.7 produced events 
(which is one of the three final states).  
For continuum process, for every pb$^{-1}$, we obtain 1.3 produced events.
\end{itemize}

From the above estimates, the integrated luminosity of the  
off resonance data sample need to
be the same order of magnitude as on $\psip$ resonance. 

According to the experience of BES data taking, the typical integrated 
luminosity at $\psip$ mass is about 8 nb$^{-1}$ per RUN;
average 25 RUNs can be taken every day, or equivalently 
0.2 pb$^{-1}$ per day. As an estimation, using these experiment
data and form factors, the time needed for events numbers 
taken at the continuum of different processes
can be worked out and are given in Table \ref{timned}.

\begin{table}[thb]
\caption{\label{timned} Events number and time needed}
\vskip 0.2 cm
\center
\begin{tabular}{cc||ccccc}  \hline \hline
\multicolumn{2}{c||}{Events number}
                     &  1  &  4  &  10  &  25  &  50   \\ \hline \hline
\multicolumn{1}{c|}{data taking}
           & $\ogpi$ & 1.3 & 5.3 &  13  &  33  &  66   \\ 
\multicolumn{1}{c|}{duration}
           & $\PP$   & 1.5 & 6.1 &  15  &  38  &  76   \\ 
\multicolumn{1}{c|}{ (day)}
           & $\rpi$  & 3.8 &15.4 &  38  &  96  &  192  \\ \hline \hline
\end{tabular}
\end{table}
In order to obtain reliable signal, and 
taking both statistical error and data collection time into consideration,
we recommend that at least 10 pb$^{-1}$ data at off $\psip$ 
peak to be taken. From the above estimates, it gets the 
statistic to match the new sample about 20 pb$^{-1}$ of  
$\psip$ taken by BES-II. 
It will serve the purpose to take
into account the contribution from one-photon continuum
in data analysis.

\section{Summary}

In this paper, we discussed extensively the properties of the 
observed cross section at resonance in $\EE$ experiment.
We studied the resonance and the electromagnetic processes.
We also pointed out the possible uncertainty due to actual 
measurement circumstance. The electromagnetic final states of
$\ogpi$ and $\PP$ are from two processes: one is
$\psip$ decay and the other the continuum contribution. Using
form factors, we found the continuum contribution is 
equivalent to the previously reported cross sections of the 
resonance decay. This indicates that the actual branching ratios 
of these electromagnetic decays might be different from 
the present reported values.

In order to obtain the correct results of $\psip$ and $\psipp$ decay, 
we suggest that at least 10~pb$^{-1}$ data be taken at the continuum region 
($E_{cm}=3.67$ GeV).

\end{document}